\documentstyle[prl,aps,epsf]{revtex}
\draft
\begin{document}
\twocolumn[\hsize\textwidth\columnwidth\hsize\csname @twocolumnfalse\endcsname

\title{Effect of FET geometry on charge ordering of transition metal oxides
} 
\author{C.J. Olson Reichhardt, C. Reichhardt, D.L. Smith, and A.R. Bishop} 
\address{ 
Theoretical Division, 
Los Alamos National Laboratory, Los Alamos, New Mexico 87545}

\date{\today}
\maketitle
\begin{abstract}
We examine the effect of an FET geometry on the charge ordering
phase diagram of transition metal oxides using numerical simulations
of a semiclassical model including long-range Coulomb fields,
resulting in nanoscale pattern formation.
We find that the phase diagram is unchanged for insulating layers
thicker than approximately twice the magnetic correlation length.
For very thin insulating layers,
the onset of a charge clump phase is shifted to lower values of the strength
of the magnetic dipolar interaction, 
and intermediate diagonal stripe and geometric phases can be
suppressed.  Our results indicate that, for sufficiently thick
insulating layers, charge injection in an FET geometry can be
used to experimentally probe the intrinsic charge ordering phases in
these materials.
\end{abstract}

%\vskip2pc
\vskip2pc]
\narrowtext

Charge ordering in doped transition metal oxides has attracted considerable
recent interest, both in theory and experiment.  Due to the competing
long-range, e.g., Coulomb, repulsion and short range 
antiferromagnetic interactions in the charge system,
a rich variety of phases can occur, including stripes \cite{Emery93,Emery97},
clumps \cite{Koulakov96,Fogler96}, and liquid crystalline electron
states \cite{Fradkin99}.  Simulation studies of 
the charge ordering phase diagram \cite{BrankoPRL,BrankoPRB}
showed transitions among four phases depending on the hole density and
the strength of a dipolar interaction
induced by the holes: a Wigner crystal at low hole densities,
a diagonal stripe phase, an intermediate geometric phase, and
a clump phase at larger dipole interaction strengths.  The behavior
of these phases is of interest not only for the charge ordered system,
but also for similarities to other pattern forming systems
with coexisting short- and long-range interactions,
including magnetic films \cite{Seul}, Langmuir monolayers,
polymers, gels, and water-oil mixtures \cite{Gelbart}.
A natural extension of this model is to 
consider charges interacting with a distortable charged membrane, which
could be relevant to active membrane systems such as ion pumps.

\begin{figure}
\center{
\epsfxsize=3.5in
\epsfbox{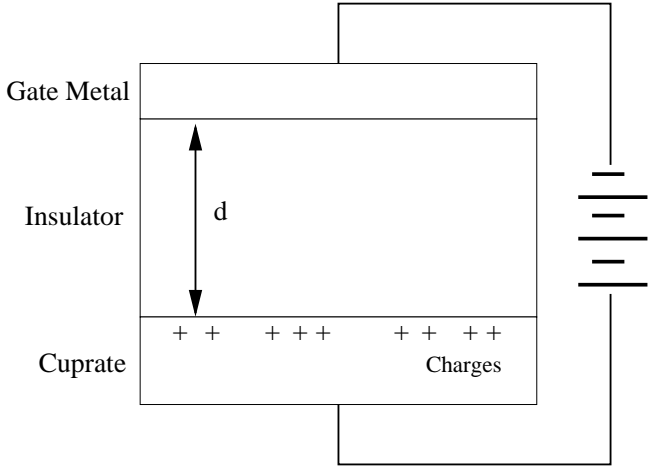}}
\caption{Schematic of the FET geometry considered here.}
\label{fig:mosfet}
\end{figure}

To experimentally probe the charge ordering phase diagram, the
hole doping of the material must be controlled.  A recently proposed
method of controllably varying the hole density is the use of a
field effect transistor
(FET) geometry to inject holes into the metal oxide plane.
The geometry is illustrated in Fig.~\ref{fig:mosfet}.
An insulating layer is deposited on top of the metal oxide, and then
a metallic gate is deposited on top of the insulator, forming a capacitive
structure.  By varying the gate voltage, holes move into or out of the
metal oxide layer, allowing the sample to be conveniently tuned to
the desired doping level.  Source and drain contacts, not shown in
the figure, can be used to probe the conductance properties of
the structure.
A potential drawback of this geometry for mapping
the phase diagram is that the holes can interact with the gate
layer, in addition to the intrinsic hole-hole interactions within the metal
oxide which lead to the charge ordered phases.  As a result, the phases
could be distorted or disrupted by the presence of the FET geometry.

To asses the effect of an FET geometry on the charge ordering phases,
we simulate a model of a 
single metal oxide layer interacting with a metallic gate 
layer that is offset by varying thicknesses of insulating material.  
We find that for a sufficiently thick insulating layer, the charge 
ordering is unaffected by the presence of the gate.  When the insulating 
layer thickness approaches twice the magnetic correlation length in the 
metal oxide, however, we find a dramatic downward shift in the onset of the 
clump phase as a function of dipole interaction strength.  
The diagonal stripe and geometric phase boundaries do not shift, but these  
phases may be suppressed by the intrusion of the clump phase as
the insulating layer is made thinner.  Our results show that, for
sufficiently thick insulating layers, the FET geometry provides a
reliable probe of the charge ordering phase behavior.

We consider a sample constructed in an FET geometry,
illustrated in Fig.~\ref{fig:mosfet}.  The metal
oxide plane is parallel to the insulating layer and also to the
metallic gate layer 
deposited on top of the insulator.  Experimentally,
a gate 
voltage is used to tune the doping level present in the 
metal oxide
layer.  We simulate this effect by directly 
varying the hole density
in our system, which is a
rectangular computational box of size $L_x \times L_y$,
with $L_x$, 
$L_y$ up to 100 unit cells in a CuO$_2$ plane.  At the
beginning of each simulation, we place the holes at random and assign
to each hole a magnetic dipole moment of constant size, but random
direction.  We find the minimum of the total potential in this
system using our efficient Monte Carlo method, described in \cite{BrankoPRB}.

Our model for the interactions between the charges
is based on the spin density wave picture of the transition
metal oxides.  Full details of the model can be found in
\cite{BrankoPRL,BrankoPRB}.
The system is doped with holes with planar 
%reattach paragraph

%FIGURE MARK

\hspace{-13pt}
density 
$\sigma_{s}$ in
the lower Mott-Hubbard band.  The interaction between the holes 
consists of both a short-range
attractive interaction, and a long-range dipolar interaction
\cite{Schrieffer,Shraiman}.  These interactions arise from the
AF bond breaking and the (dipolar in range) spiral distortions
of the AF background \cite{Shraiman,Frenkel}.
The system is magnetically disordered at finite temperature, and is
characterized by a finite magnetic correlation length $\xi$
\cite{Chakravarty}, which is determined from neutron
scattering measurements \cite{Birgeneau}.  We take $\xi=3.8/\sqrt{n}$ \AA .
The doping level $n$ is defined as the hole density measured
in units of the cuprate lattice spacing; thus $n=1\%$ corresponds
to 1 hole per $100a^{2}$, where $a\approx 3.8$\AA.
We take the long-ranged Coulomb interaction between the holes to be
unscreened, as appropriate at low doping where $r_s=r_0/a_0$ is
very large: $r_0$ is the mean interparticle distance, $a_0$ is
the Bohr radius, and $r_s \approx 8$.  In our model, the effective interaction
between two holes a distance ${\bf r}$ apart
in a single metal oxide plane is then of the form
\begin{equation}
V({\bf r})=\frac{q^2}{r} - B \cos(2\theta - \phi_1 - \phi_2)e^{-r/\xi} , 
\end{equation}
where $q$ is the hole charge, $\theta$ is the angle between ${\bf r}$ and
a fixed axis, and $\phi_{1,2}$ are the angles of the magnetic dipoles
relative to the same fixed axis, which we assume can take an arbitrary
value.  $B$ is the strength of the magnetic 
dipolar interaction
[$B \approx U/(2\pi \xi^{2})$], which, in real materials, should
be of order $\sim 1$eV.  We have introduced $B \sim B_{xy}/l^{2}$,
where $l$ is some appropriate average length ,
$a<l<\xi$, in order to avoid the unphysical divergence of the
dipolar part of the interaction, while keeping the necessary
symmetry of the interaction.
We have assumed that ${\bf r}$ can be
relaxed from a crystal lattice position to an arbitrary (continuous)
value.
The results presented here are for a system with 196 holes
with size ranging from 307.15\AA $\times$ 307.15\AA \ (for $n=3\%$) to
137.4\AA $\times$ 137.4\AA \ (for $n=15\%$).
The value of $\xi$ ranges from 21.9\AA \ for $n=3\%$ to
9.8\AA \ for $n=15\%$.  The sample is periodic in the x-y 
%reattach paragraph

%FIGURE MARK
\begin{figure}
\center{
\epsfxsize=3.5in
\epsfbox{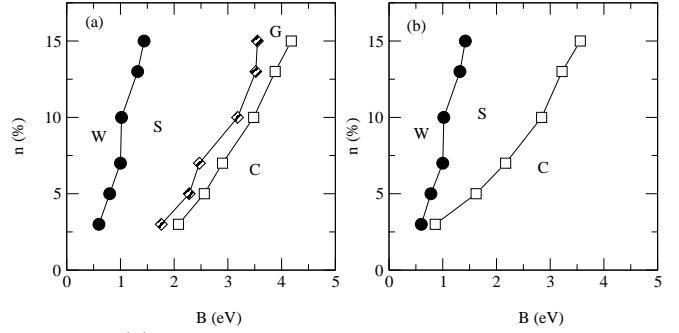}}
\caption{(a) Phase diagram as a function of the hole density 
$n$ and the strength
of the magnetic dipolar interaction $B$, for a sample without the
FET interaction term.  ``W'' is the Wigner crystal phase,
``S'' is the diagonal stripe phase, ``G'' is the geometric phase,
and ``C'' is the clump phase.  (b) Phase diagram for a sample with the
same parameters but with the
FET interaction added at an insulator thickness of 14\AA, showing the
downward shift in $B$ of the clump phase ``C'' and the suppression of the
geometric phase ``G''.
}
\label{fig:phase}
\end{figure}

\hspace{-13pt}
plane only
\cite{Jensen}.  

In the FET geometry, a metal oxide channel is created,
then an insulating layer of thickness $d$ is deposited on top
of the channel, and finally a layer of metal is deposited to serve
as a gate.  The interactions between
holes in the metal oxide layer is altered by the presence of image
charges in the gate layer,
\begin{equation}
\nabla \cdot {\bf D}=4\pi\rho 
= 4\pi e \sum_{i}
[\delta(r-(r_{||i}+d{\hat z})) - \delta(r-(r_{||i}-d{\hat z}))] ,
\end{equation}
giving the Coulomb energy between charges as
\begin{equation}
E=\frac{e^2}{2\epsilon}\sum_{ij}^{\prime}\left[
\frac{1}{|r_{||j}-r_{||i}|} - \frac{1}{|r_{||j}-r_{||i}+2d{\hat z}|}\right] .
\end{equation}
We modify the Coulomb interaction between the holes in our system 
to this form.  This introduces a new length scale $2d$. 
As a comparison, we also run simulations with the
unmodified Coulomb interaction, representing a bare metal oxide plane
without the FET gating.

In the absence of the FET interaction, we find a phase diagram
consistent with that observed in Ref. \cite{BrankoPRL}, as illustrated
in Fig.~\ref{fig:phase}(a).  
For thick insulating layers, 
when the FET interaction is included, the locations of the phases
are not affected and we obtain
the same phase diagram, as shown in Fig.~\ref{fig:phase}(a).  
As we decrease the thickness of the insulating layer, we find a
critical thickness $d_c$ below which the phase boundaries begin
to shift.  Fig.~\ref{fig:phase}(b) 
illustrates the phase diagram for the same system in
Fig.~\ref{fig:phase}(a) 
but with the FET term added and with an insulating layer
of thickness $d<d_c$, $d=14$ \AA.  The onset of the diagonal stripe phase 
``S'' is unaffected, but there is a 
large shift downward in $B$ of the onset
of the clump phase ``C,'' $B_C$.  The size of the downward shift $\Delta B$
increases as the 
%reattach paragraph

%FIGURE MARK
\begin{figure}
\center{
\epsfxsize=3.5in
\epsfbox{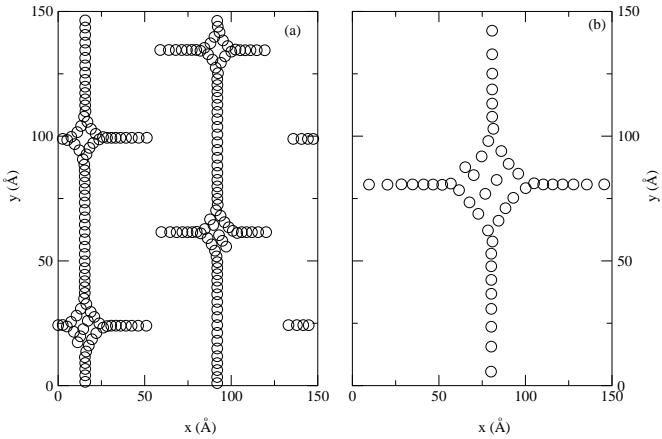}}
\caption{The clump phase at two different hole densities.
(a) $n=13\%$, $\xi=10.54$\AA.  (b) $n=3\%$, $\xi=21.9$\AA.
}
\label{fig:clump}
\end{figure}

\hspace{-13pt}
hole density $n$ decreases.  The geometric phase that
was present without the FET term is now suppressed completely.

The interplay of the correlation length and the
insulator thickness affects the onset
of the clump phase because only the clump phase possesses a characteristic
length 
scale of order $2\xi$.
This is illustrated in Fig.~\ref{fig:clump}, which shows 
the clump
structure at $n=3\%$ and $n=13\%$. 
When the length 
scale of the insulating layer is similar to the length scale of
the clump structure, the interaction with the metal gate above the
insulating layer becomes comparable with the interaction with
neighboring clumps, and the transition to the clump phase is enhanced.

\begin{figure}
\center{
\epsfxsize=3.5in
\epsfbox{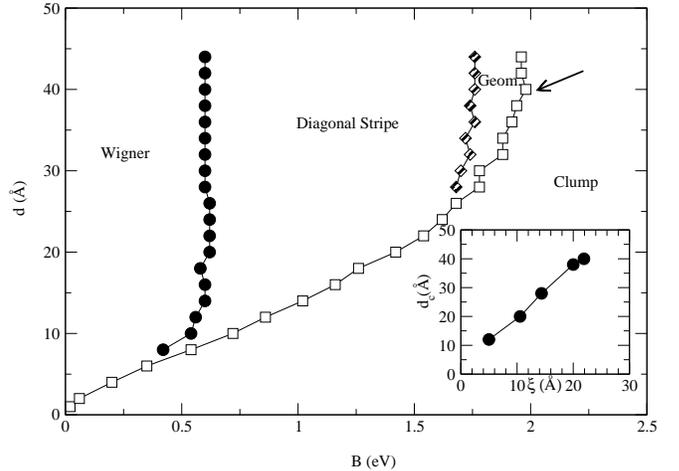}}
\caption{
Phase diagram as a function of the thickness of the insulating
layer $d$ and the strength of the magnetic dipolar interaction $B$, for
a sample with the FET interaction term and with hole density $n=3\%$.
Inset: Dependence of the cutoff thickness $d_c$ on the magnetic correlation
length $\xi$.
}
\label{fig:dz}
\end{figure}

The effectiveness of the FET term extends only to 
insulator thicknesses $d$ that
are approximately twice the magnetic correlation 
length, $d \lesssim 2\xi$.  This is illustrated
in Fig.~\ref{fig:dz} for a sample with $n=3$\% and varying insulator
thicknesses $d.$  The arrow indicates the saturation of the clump 
phase onset $B_C$ at 
$d_c$ to the value of $B=B_{C}^{0}$ observed in the absence of the FET
term.  The other phase boundaries, Wigner to stripe
and stripe to geometric, do not shift with $d$ but instead the intermediate
phases are suppressed when $B_C$ moves below the
onsets of these phases.  To test whether the magnetic correlation length 
is responsible for the shift in the phases, we considered a
system at $n=13\%$, which normally has a correlation length of
$\xi = 10.54$ \AA, and artificially changed the screening length to 
$\xi = 20$ \AA \ for
one series of runs, and to $\xi = 5$ \AA \ for a second series.  In the
inset of Fig.~\ref{fig:dz}, we show the cutoff insulator thickness 
$d_c$ beyond which
the FET term has no effect as a function of correlation length
$\xi$, both
for the normal screening lengths and for our two artificially changed
screening lengths.  We find a linear dependence for the
cutoff thickness on the correlation length, $d_c \approx 1.7 \xi$.

In conclusion, we find that the FET geometry does not affect
the clump ordering phases unless the insulating 
layer is thin enough,
namely less than
approximately twice the magnetic correlation length.
For thin 
%reattach paragraph

%FIGURE MARK

\hspace{-13pt}
insulating layers the onset of the clump phase is enhanced.
This suggests that the FET geometry with a sufficiently thick
insulating layer is suitable for studying the effects of hole
concentration on the charge ordering phases because
the presence of the FET does not alter the phase structure.
Alternatively, FET geometries deliberately created with thin
insulating layers can be used to probe the clump phase at higher
hole densities, where the underlying value of $B$ may preclude reaching
the clump state.  

This work was supported by the U.S. Department of Energy
under Contract No. W-7405-ENG-36.

\end{document}